\documentclass[twocolumn]{revtex4-1}

\usepackage{graphicx}
\usepackage{csquotes}
\usepackage[utf8]{inputenc}
\usepackage[T1]{fontenc}
\usepackage{amsmath} 
\usepackage{amsthm} 
\usepackage{esint}
\usepackage{esvect}
\usepackage{esint}

\begin{document}


\title{Ultrashort Magnetic Impulses Driven by Coherent Control with Vector Beams}


\author{S. Sederberg$^{1}$}
\author{F. Kong$^{1}$}
\author{P. B. Corkum$^{1}$}
\affiliation{$^1$Joint Attosecond Science Laboratory, University of Ottawa and National Research Council of Canada, 100 Sussex Drive, Ottawa K1N 5A2, Ontario, Canada}


\date{\today}

\begin{abstract}
\setlength\parindent{0em}
We introduce a new technique for the generation of magnetic impulses. This technique is based on coherent control of electrical currents using cylindrical laser beams with azimuthal polarization. When used to ionize a medium, in this case atomic hydrogen is considered, an azimuthal current impulse is driven. The spatial distribution of this current bears close resemblance to that of a solenoid, and produces a magnetic field impulse. The excitation and relaxation dynamics of this current temporally confine the resulting magnetic field to a Tesla-scale, terahertz bandwidth impulse. Importantly, the magnetic fields are spatially isolated from electric fields. This all-optical approach will enable ultrafast time-domain spectroscopy of magnetic phenomena. 
\end{abstract}

\pacs{}

\maketitle


\setlength\parindent{0em}
Magnetic fields play a central role in many areas of physics, including magnetic materials and devices, superconductivity, electron spin manipulation, magnetic-field-induced phase transitions, quantum and topological systems, quantum critical points, plasma physics, and plasma confinement for nuclear fusion. The Biot-Savart law, formalized in 1820, provides a comprehensive classical description of static magnetic fields generated by charged particles in motion. Confining current to a carefully chosen path, typically a variation of a solenoid, enables strong, uniform magnetic fields to be generated in the region enclosed by the circulating current. The magnetic field present at the center of an ideal solenoid is determined by the simple expression: $B=\mu nI$, where $\mu$ is the permeability of the core medium, $n=N/L$ is the number of wire turns per unit length, and $I$ is the current flowing through the wire. 

\setlength\parindent{1em}

Electrical conductors have been quintessential to solenoids, and technical challenges associated with their resistance have hindered advancement of the amplitude and bandwidth of magnetic field sources. For example, the latency associated with the electrical circuitry comprising conventional electromagnets prevents temporal resolution of sub-picosecond magnetic-field-induced dynamics and as a result, the microscopic origins responsible for the magnetic response of a sample must be inferred from static measurements.

Wires imposed comparable challenges on the study of electric field phenomena in solid state systems. Phase-locked terahertz electromagnetic pulses emerged as the first metrological tool capable of resolving electronic phenomena on timescales rendered prohibitive by electrical transmission lines \cite{Ulbricht2011,Auston1984,Wu1995,Leitenstorfer1999}. Using electric fields derived from short laser pulses not only enabled picosecond and femtosecond temporal resolution, but also dramatically reduced avalanche breakdown and thermal damage, enabling new regimes of electric field strength to be explored. Over the last two decades, our ability to interrogate matter using short electromagnetic pulses has expanded to the near-infrared and optical spectral regions, enabling electron dynamics to be probed on their natural, attosecond timescale \cite{Goulielmakis2004,Kim2013,Keiber2016,Sommer2016}. Complementary magnetic field sources would permit insight into magnetic field phenomena on unprecedented timescales.


Relaxing the requirement for a solenoid fabricated from wires, one can envisage a solenoid composed of a continuous conductive sheet. In this case, the magnetic field at the center of the solenoid would be expressed as: $B=\mu I_L$, where $I_L$ is the current per unit length circulating through the sheet. When a sufficiently intense laser pulse is incident on a medium, such as a gas, it will ionize the medium and drive the electron population in its optical fields. 

In this Letter, we show that the excitation of a gas by intense laser pulses can produce a current distribution that closely resembles that from a conventional solenoid. Two cylindrical vector beams with azimuthal polarization are used to ionize atomic hydrogen and drive an electrical current circulating the longitudinal axis of the beam, closely resembling current flowing in a solenoid. The direction and amplitude of this current, and of the induced magnetic field, can be precisely controlled by adjusting the relative phase between the two light waves. These calculations demonstrate a conceptually simple technique for exciting and probing samples with isolated magnetic impulses of picosecond duration, providing a new approach for studying magnetic phenomena on picosecond and femtosecond timescales. We demonstrate a simple approach to scale the magnetic field amplitude and investigate the influence of back electromotive force as magnetic fields on the Tesla scale are excited on a femtosecond timescale. These proof-of-principle calculations demonstrate the feasability of scaling the magnetic amplitude to at least $0.42$T, and approaches for further scaling are proposed. 


  \begin{figure}
 \includegraphics[width=0.50\textwidth]{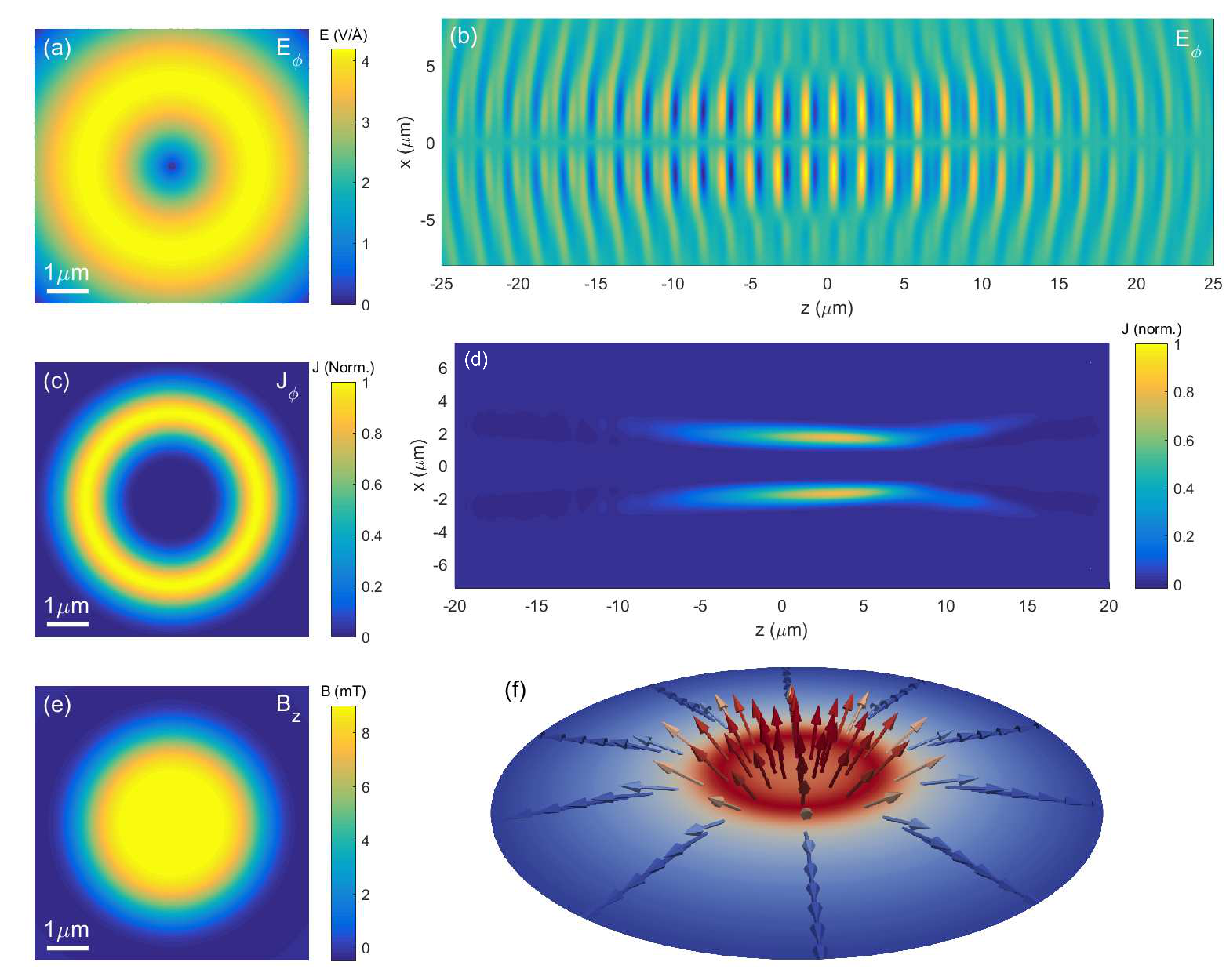}
 \caption{(a) The azimuthal electric field component of the laser beam. (b) A snapshot of the electric field from a cylindrical beam with azimuthal polarization propagating through the beam waist. The laser fields produce an azimuthal current density bearing close resemblance to current flowing in a solenoid, which is shown for a tranverse cross-section in (c) and a longitudinal cross-section in (d). (e) This current density produces a uniform magnetic field that is enclosed by the solenoidal current. (f) Magnetic vector field including both $B_z$ and $B_r$ components. \label{fig1}}
 \end{figure}

The aforementioned currents have been routinely generated in solids and gases through coherent control. In general, coherent control is an optical technique whereby a laser pulse with angular frequency, $\omega,$ is applied to a system simultaneously and collinearly with its second harmonic, $2\omega$ \cite{Miller1980, Jackson1983, Chen1990, Yin1992}. The system can be driven into an excited state by each of the two pulses independently. Applying the two pulses coherently and adjusting their relative phase enables control of quantum interference between the two processes, which can be detected from any observable related to the excited state. In atomic systems, the two laser pulses are commonly used to ionize an atom, and their relative phase controls the direction and energy of the emitted photoelectrons. In solid state systems, the relative phase can be used to affect asymmetry on the momentum distribution of a photoexcited conduction band population, driving a current through the material \cite{Dupont1995,Hache1998}. The ability to drive currents using optical fields, without the need for electrical conductors, holds potential technological and metrological applications, and plays a central role in the work presented here.

The increasing availability of structured laser modes provides the possibility to transfer intricate features from the laser beam mode profile, polarization, and orbital angular momentum to the collective electron motion through the coherent control process. As a specific case, we consider a cylindrical vector beam with azimuthal polarization, which is depicted in Fig. 1(a) \cite{Zhan2009}. We numerically investigate strong field ionization in atomic hydrogen, where the tunnelling ionization rate can be expressed as:
\begin{equation}
w(t) = 4\omega_0\Bigg(\frac{E_a}{E(t)}\Bigg)\textrm{exp}\Bigg[-\frac2{3}\Bigg(\frac{E_a}{E(t)}\Bigg)\Bigg],
\end{equation}
where $\omega_0=me^4/\hbar^3$ is the atomic unit of frequency, $E_a=m^2e^5/\hbar^4$ is the atomic unit of the electric field, and $E(t)$ is the time-varying electric field incident on the medium. Ionization and classical trajectories are calculated for a laser pulse at $\lambda=1800$nm and its second harmonic at $\lambda=900$nm with peak electric field strengths of $E_{p,\omega}=3.0\textrm{V/\AA}  (I_{p,\omega}=1.2\times10^{14}$W/cm$^2)$ and $E_{p,2\omega}=1.5\textrm{V/\AA}  (I_{p,2\omega}=0.6\times10^{14}$W/cm$^2)$, respectively, and identical pulse durations of $\tau_p=18$fs.

We simulate this physical scenario using an approach that is closely related to particle in cell calculations. We propagate the two pulses through the Rayleigh length of a Gaussian focusing geometry (beam waist, $w_0=3\mu$m) using three-dimensional finite-difference time-domain simulations in cylindrical coordinates, as depicted in Fig. 1(b) \cite{Taflove1995}. We spatially discretize the simulation space with $\Delta r=\Delta z=40$nm and $\Delta\theta=\pi/150$, and use a timestep of $\Delta t=76$as. At each time-step, the ionization rate is evaluated at every mesh point in the simulation space, electric fields are interpolated to the position of each existing electron, and all existing electron trajectories are updated. The existing trajectories are distributed into their nearest mesh points to produce an effective current density mapping, which is included in the next time-step of Maxwell's equations for self-consistency. In particular, this ensures that the back-action of the time-dependent magnetic fields (i.e. back electromotive force) on the excited currents is accounted for.
 
In the vicinity of the beam waist, the extremely nonlinear photoionization process spatially gates the ionization process affected by the azimuthal beams, confining the excitation of an azimuthal current to a thin circular loop in the transverse plane, as shown in Fig. 1(c). We emphasize that this is an impulsively excited current, whereby electrons at each azimuthal position, $\delta\theta$, are briefly imparted with momentum in the azimuthal direction, producing a pulsed collective current excitation that resembles a solenoidal current. This current induces and encompasses a uniform magnetic field, and a tranverse slice of the $B_z$ field is shown in Fig. 1(e). The full-vectorial magnetic field distribution is depicted in Fig. 1(f), and demonstrates that the field is predominantly longitudinal.
 
\begin{figure}
 \includegraphics[width=0.45\textwidth]{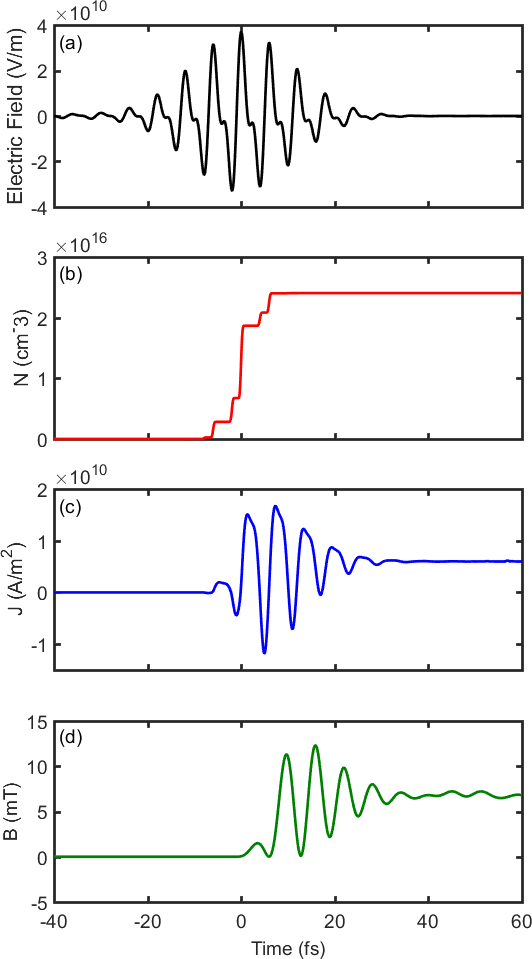}
 \caption{Temporal dynamics. (a) An exemplary $\omega/2\omega$ electric field waveform at the beam waist. (b) Growth of the ionized population density when this field is incident on a hydrogen gas target with a density, $N_0=10^{17}$cm$^{-3}$. (c) The current density dynamics that occur as the exicted population interacts with the remaineder of the laser pulse. (d) The magnetic field produced by this current density. \label{fig2}}
 \end{figure}
 
The temporal waveform of the electric field excitation at the beam waist is shown in Fig. 2(a). With the highest ionization rate occurring at the peak of the laser pulse, shown in Fig. 2(b), the current density dynamic is confined mainly to the latter half of the laser pulse, as shown in Fig. 2(c). A magnetic field is induced by this current, and its dynamic build-up is presented in Fig. 2(d).This magnetic field will exist for as long as the azimuthal current maintains coherence. After the laser pulses have left the ionized atoms, a collective response arising from the screening potential of each charge carrier will develop over a timescale on the order of the inverse of the plasma frequency \cite{Huber2001}. For an ionization density of $n=10^{17}\textrm{cm}^{-3}$, we calculate a plasma frequency, $\omega_p=\sqrt{\frac{ne^2}{m_e\varepsilon_0}}=2.8$THz, and a corresponding time constant on the order of $\tau_r=350$fs. The excitation ($\tau_p$) and relaxation ($\tau_r$) temporally confine the current to a unipolar transient, which radiates a single-cycle terahertz pulse with a space-time structure closely resembling a flying electromagnetic doughnut \cite{Hellwarth1996}. This teraherz pulse can easily be spectrally isolated from the near-infrared excitation pulses using a long-pass filter.

Adjusting the relative phase between the $\omega$ and $2\omega$ laser pulses imposes coherent control on the excited current density and the resulting magnetic fields. Figure 3(a) shows the peak magnetic field at the beam waist as the relative phase between the two pulses is adjusted. Figures 3(b)-(e) show the magnetic field distribution in the longitudinal plane of the focusing geometry for $\Delta\phi = \pi/8, 5\pi/8, 9\pi/8,$ and $13\pi/8,$ respectively. When the relative phase is adjusted to produce a magnetic field maximum in the beam waist, the magnetic field distribution closely resembles that from a solenoid. Conversely, when the relative phase is adjusted for a minimum at the beam waist, the magnetic field changes sign due to the Gouy phase shift of the focused fields.

 \begin{figure}
 \includegraphics[width=0.50\textwidth]{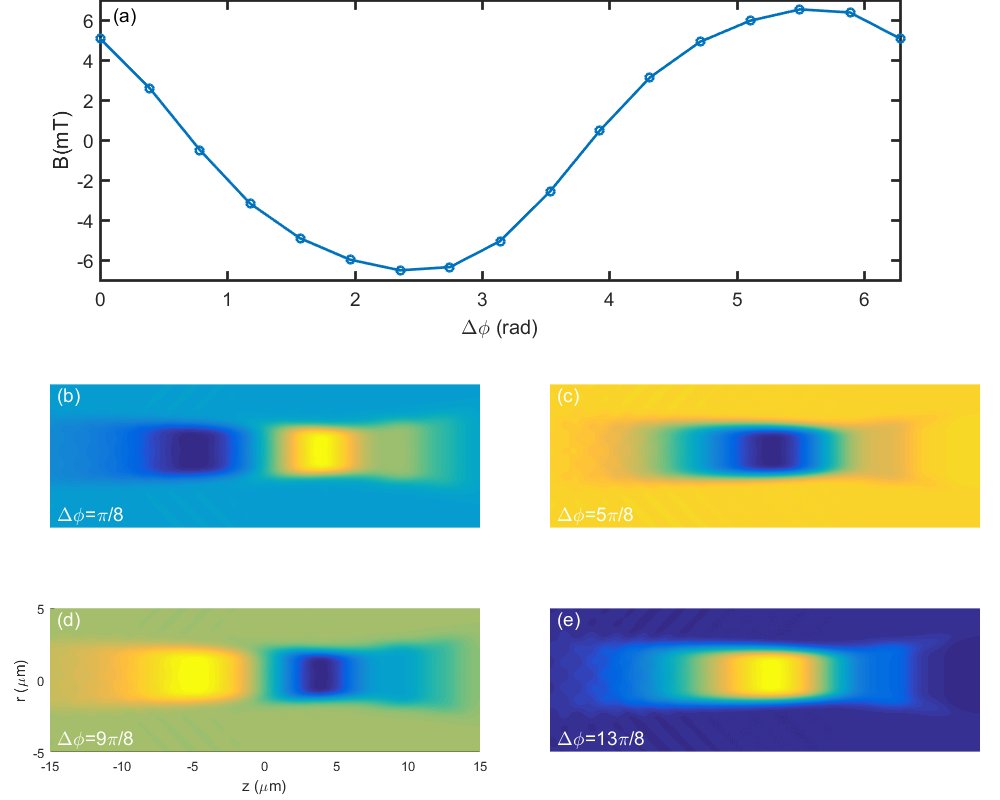}
 \caption{Coherent control of magnetic fields. (a) Adjusting the phase offset between the $\omega$ and $2\omega$ beams enables precise control of the direction and amplitude of the generated magnetic field. Longitudinal slices of the magnetic fields generated in the focusing geometry are shown for (b) $\Delta\phi=\pi/8$, (c) $\Delta\phi=5\pi/8$, (d) $\Delta\phi=9\pi/8$, and (e) $\Delta\phi=13\pi/8$. \label{fig3}}
 \end{figure}
 
\begin{figure}
 \includegraphics[width=0.50\textwidth]{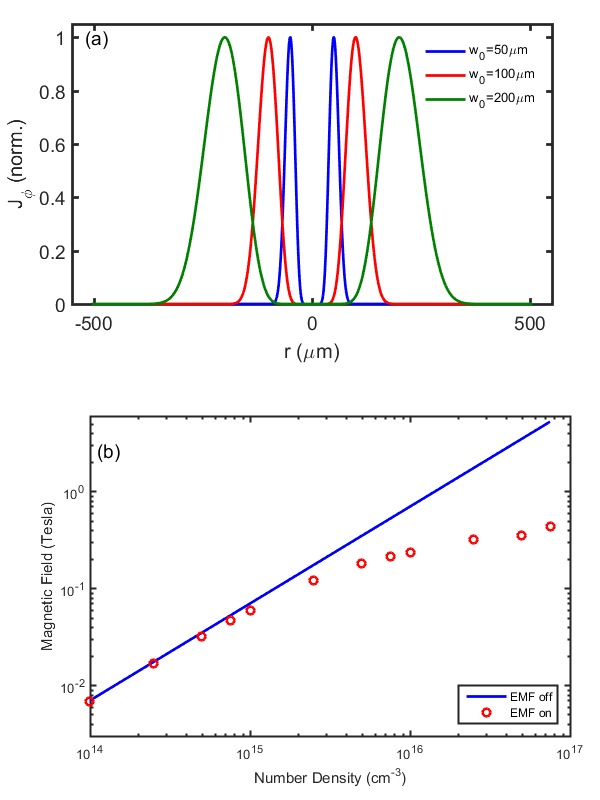}
 \caption{(a) Increasing the beam waist, $w_0$ of the azimuthal laser beam increases the radial extent of the current density that is produced at the focus, thereby increasing the current per unit length flowing through the solenoid. (b) Magnetic field for a constant beam waist, $w_0=200\mu$m, and varying gas density. Two simulations are performed for each gas density: one neglecting back electromotive force and the other including it. \label{fig4}}
\end{figure}

Notably, the spatial distribution of the current density shown in Figs. 1(c)-(d) bears strong resemblance to that in a solenoid. The Rayleigh length of a focused laser beam increases quadratically with the beam waist, i.e. $z_R=\pi w_0^2/\lambda$, and hence, for loose focusing conditions the current density distribution converges to that of an ideal solenoid. Increasing $w_0$ while keeping the peak electric field constant (i.e. by increasing the laser pulse energy) produces a larger solenoidal current excitation per unit length, as depicted in Fig. 4(a). Due to the independence of the solenoidal magnetic field on the solenoid radius, this provides a simple and convenient means to scale the magnetic field amplitude. With the availability of energetic femtosecond laser pulses, a crucial question that must be addressed is, to what magnetic field amplitude may this scheme be scaled before the electric field induced by a rapidly excited magnetic field is sufficient to suppress the excited current density and resulting magnetic field? In other words, when does the opposing electric field amplitude excited by a time-dependent magnetic field as determined by Faraday's Law,

\begin{equation}
\oint \vv{\mathbf{E}}\cdot\textrm{d}\vv{\mathbf{l}}=-\int\frac{\partial\vv{\mathbf{B}}}{\partial t}\cdot\textrm{d}\vv{\mathbf{A}},
\end{equation}
become significant compared to the electric fields of the laser pulse?

To investigate the influence of this opposing electric field on the excited electron trajectories, we increase the beam waist to $w_0=200\mu$m which, according a simple application of the solenoidal relationship ($B=\mu I_L$) could generate Tesla-scale magnetic fields. We consider a fundamental laser pulse at $\lambda=4000$nm with intensity $I=1.2\times10^{14}$W/cm$^2$, and a corresponding second harmonic at $\lambda=2000$nm with intensity $I=3\times10^{13}$W/cm$^2$. Each pulse has a duration of $\tau=50$fs. Keeping the excitation parameters fixed, we increment the gas density from $n=1\times 10^{14}$cm$^{-3}$ to $n=7.5\times 10^{16}$cm$^{-3}$ and perform two simulations for each gas density: one with back electromotive force included and the other neglecting this effect. We note that in the absence of back electromotive force, one would expect linear scaling of the current density and magnetic field with respect to the gas density. The result of these simulations is shown in Fig. 4(b). In close agreement with the solenoidal relationship ($B=\mu I_L$), magnetic field amplitudes up to $5.2$T are observed in the absence of back electromotive force. However, the simulations including back electromotive force demonstrate considerable suppression of the excited magnetic field amplitude as the gas density is increased beyond $10^{15}$cm$^{-3}$, and the highest gas density produces a magnetic field amplitude of $0.43$T. Although back electromotive force appears to be the dominant saturation effect in this magnetic field scaling scheme, these calculations demonstrate the prospect for all-optical generation of spatially isolated magnetic fields with amplitudes that have technological and metrological significance.


We note two routes that could potentially enable further scaling of the magnetic field amplitude. The higher ionization potential of helium ($I_p=24.6$eV) compared to hydrogen ($I_p=13.6$eV) would permit higher laser intensities incident on the gas, driving the photoelectrons to a greater final velocity. Alternatively, we envision that a variation of this experiment incorporating a circularly polarised beam, which has been shown to produce high-energy above threshold ionization electrons \cite{Corkum1989}, and an orbital angular momentum beam could enable an order-of-magnitude increase in the magnetic field amplitude.
 
In conclusion, we have proposed an all-optical scheme to generate magnetic field impulses. Solenoidal electrical currents are generated through coherent control of strong field ionization, where an azimuthal electric field drives an azimuthal current. The amplitude and direction of this current can be precisely controlled by changing the relative phase between the $\omega$ and $2\omega$ light waves. Magnetic fields approaching the Tesla scale can be produced by using loose focusing conditions and mid-infrared laser pulses. While the modulation bandwidth of magnetic fields in integrated magneto-optical circuits has been limited to the GHz range imposed by electrical circuitry, this provides a route to magneto-optical modulation beyond THz frequencies. 

Perhaps the most intriguing property of the calculated magnetic fields is their spatial localization. The longitudinal polarization of the generated magnetic field allows the magnetic field direction introduced to a system, such as an electron spin, to be modulated simply by changing its angle of incidence on the region of interest, providing remarkable flexibility for spin logic devices. The purely longitudinal magnetic field is naturally spatially isolated from coupled electric fields, granting access to metrology and technology based purely on magnetic fields.The introduction of a short, intense magnetic impulse in the form of an electromagnetic wave would also enable optical synchronization with sophisticated detection schemes, such as magneto-optical sampling \cite{Riordan1997}, high-harmonic magnetic circular dichroism spectroscopy \cite{Willems2015}, or spin-polarized attosecond electron diffraction \cite{Kealhofer2016, Morimoto2018}, enabling temporal resolution of magnetic-field-induced microscopic dynamics. Magnetic field metrology based on ultrafast optical techniques will enable new frontiers in magnetism, particularly scaling to higher magnetic fields and resolution of femtosecond and attosecond magnetization dynamics. 

\vspace{2mm}

This research was supported by the Natural Sciences and Engineering Research Council of Canada (NSERC) Discovery Grant Program, the Canada Research Chairs program, and the United States Defense Advanced Research Projects Agency (``Topological Excitations in Electronics (TEE)'', agreement \#D18AC00011).

\end{document}